\documentclass[pre,twocolumn,a4paper,showpacs,floatfix]{revtex4}
\usepackage{amsfonts}
\usepackage{amsmath}
\usepackage{amssymb}
\usepackage{textcomp}
\usepackage{bm}
\usepackage[dvips]{graphicx}
\usepackage[latin1]{inputenc}

\def\bal#1\eal{\begin{align}#1\end{align}}
\renewcommand{\vec}[1]{\bm{#1}}
\newcommand{\beq}{\begin{equation}}
\newcommand{\eeq}{\end{equation}}
\newcommand{\bsub}{\begin{subequations}}
\newcommand{\esub}{\end{subequations}}

\newcommand{\ueo}{u_\mathrm{eo}}
\newcommand{\Pmech}{P_\mathrm{mech}}
\newcommand{\Pkin}{P_\mathrm{kin}}
\newcommand{\Ptot}{P_\mathrm{tot}}

\newcommand{\dm}{\mathrm{d}}
\newcommand{\pp}{\partial}
\newcommand{\nablabf}{\bm{\nabla}}
\newcommand{\rhoe}{\rho_\mathrm{el}}
\newcommand{\lamD}{\lambda_\mathrm{D}}
\newcommand{\tauD}{\tau_\mathrm{D}}
\newcommand{\Ekin}{E_\mathrm{kin}}
\newcommand{\EkinLin}{E^{\mathrm{LS}}_\mathrm{kin}}
\newcommand{\EkinNlin}{E^{\mathrm{NLS}}_\mathrm{kin}}
\newcommand{\EkinFull}{E^{\mathrm{FN}}_\mathrm{kin}}
\newcommand{\kBT}{k_\mathrm{B}T}

\newcommand{\SIm}{\textrm{m}}

\newcommand{\SImum}{\textrm{\textmu{}m}}
\newcommand{\SImps}{\SIm\,\SIs^{-1}}

\newcommand{\SIs}{\textrm{s}}

\begin{document}

\date{23 February 2008}

\title{A numerical analysis of finite Debye-length effects in induced-charge electro-osmosis}

\author{
 Misha Marie Gregersen$^1$,
 Mathias Bækbo Andersen$^1$,
 Gaurav~Soni$^2$,
 Carl Meinhart$^2$, and
 Henrik Bruus$^1$}

\affiliation{%
$^1$Department of Micro- and Nanotechnology, Technical University of
Denmark\\ DTU Nanotech Building 345 East, DK-2800
Kongens Lyngby, Denmark\\
$^2$Department of Mechanical Engineering, University of
California\\
Engineering II Building, Santa Barbara, CA 93106, USA}

\begin{abstract}

For a microchamber filled with a binary electrolyte and containing a flat
un-biased center electrode at one wall, we employ three numerical models to
study the strength of the resulting induced-charge electro-osmotic (ICEO)
flow rolls: (\textit{i}) a full nonlinear continuum model resolving the
double layer, (\textit{ii}) a linear slip-velocity model not resolving the
double layer and without tangential charge transport inside this layer, and
(\textit{iii}) a nonlinear slip-velocity model extending the linear model
by including the tangential charge transport inside the double layer.  We
show that compared to the full model, the slip-velocity models
significantly overestimate the ICEO flow. This provides a partial
explanation of the quantitative discrepancy between observed and calculated
ICEO velocities reported in the literature. The discrepancy increases
significantly for increasing Debye length relative to the electrode size,
i.e.\ for nanofluidic systems. However, even for electrode dimensions in
the micrometer range, the discrepancies in velocity due to the finite Debye
length can be more than 10\% for an electrode of zero height and more than
100\% for electrode heights comparable to the Debye length.


\end{abstract}

\maketitle

\section{Introduction\label{sec:Introduction}}
Within the last decade the interest in electrokinetic phenomena in general
and induced-charge electro-osmosis (ICEO) in particular has increased
significantly as the field of lab-on-a-chip technology has developed.
Previously, the research in ICEO has primarily been conducted in the
context of colloids, where experimental and theoretical studies have been
carried out on the electric double layer and induced dipole moments around
spheres in electric fields, as reviewed by Dukhin~\cite{Dukhin1993} and
Murtsovkin~\cite{Murtsovkin1996}. In microfluidic systems,
electrokinetically driven fluid motion has been used for fluid
manipulation, e.g. mixing and pumping. From a microfabrication perspective
planar electrodes are easy to fabricate and relatively easy to integrate in
existing systems. For this reason much research has been focused on the
motion of fluids above planar electrodes. AC electrokinetic micropumps
based on AC electroosmosis (ACEO) have been thoroughly investigated as a
possible pumping and mixing device. Experimental observations and
theoretical models were initially reported around year 2000
\cite{Ajdari2000,Gonzales2000,Brown2000,Green2002}, and further
investigations and theoretical extensions of the models have been published
by numerous groups since
\cite{Lastochkin2004,Debesset2004,Studer2004,Cahill2004,Olesen2006,Gregersen2007}.
Recently, ICEO flows around inert, polarizable objects have been observed
and investigated theoretically
\cite{Squires2004,Levitan2005,Squires2006,Harnett2008,Khair2008,Gregersen2009}.
For a thorough historical review of research leading up to these results,
we refer the reader to Squires~\emph{et al.} \cite{Squires2004} and
references therein.

In spite of the growing interest in the literature not all aspects of the
flow-generating mechanisms have been explained so far. While qualitative
agreement is seen between theory and experiment, quantitative agreement is
often lacking as reported by Gregersen \textit{et
al.}~\cite{Gregersen2007}, Harnett \textit{et al.}~\cite{Harnett2008}, and
Soni \textit{et al.}~\cite{Soni2007}. In the present work we seek to
illuminate some of the possible reasons underlying these observed
discrepancies.

ICEO flow is generated when an external electric field polarizes an
object in an electrolytic solution. Counter ions in the electrolyte
screen out the induced dipole, having a potential difference $\zeta$
relative to the bulk electrolyte, by forming an electric double
layer of width $\lamD$ at the surface of the object. The ions in the
diffuse part of the double layer then electromigrate in the external
electric field and drag the entire liquid by viscous forces. At the
outer surface of the double layer a resulting effective slip
velocity $\vec{v}_\mathrm{slip}$ is thus established. Many numerical
models of ICEO problems exploit this characteristic by applying the
so-called Helmholtz--Smoluchowski slip condition on the velocity
field at the electrode surface \cite{Probstein1994,Bruus2008}.
Generally, the slip-condition based model remains valid as long as
\begin{equation}\label{eq:HS_slip_cond}
\frac{\lamD}{a_c}\exp\bigg(\frac{Ze\zeta}{2\kBT}\bigg)\ll 1,
\end{equation}
where $\kBT/(Ze)$ is the thermal voltage and $a_c$ denotes the
radius of curvature of the surface \cite{Squires2004}. The
slip-velocity condition may be applied when the double layer is
infinitely thin compared to the geometrical length scale of the
object, however, for planar electrodes, condition
(\ref{eq:HS_slip_cond}) is not well defined. In the present work we
investigate to what extent the slip condition remains valid.

Squires~\emph{et al.} \cite{Squires2004} have presented an analytical
solution to the ICEO flow problem around a metallic cylinder with radius
$a_c$ using a linear slip-velocity model in the two dimensional plane
perpendicular to the cylinder axis. In this model with its infinitely thin
double layer, the surrounding electrolyte is charge neutral, and hence the
strength of the ICEO flow can be defined solely in terms of the
hydrodynamic stress tensor $\vec{\sigma}$, as the mechanical power $\Pmech
= \oint_{|\vec{r}|=a_c}\hat{\vec{n}}\cdot\vec{\sigma}\cdot
\vec{v}_\mathrm{slip}\dm a$ exerted on the electrolyte by the tangential
slip-velocity $\vec{v}_\mathrm{slip} = \ueo\hat{\vec{t}}$, where
$\hat{\vec{n}}$ and $\hat{\vec{t}}$ is the normal and tangential vector to
the cylinder surface, respectively. In steady flow, this power is equal to
the total kinetic energy dissipation $\Pkin =
\frac{1}{2}\eta\int_{a_c<|\vec{r}|} (\pp_iv_j+\pp_jv_i)^2\dm\vec{r}$ of the
resulting quadrupolar velocity field in the electrolyte.

When comparing the results for the strength of the ICEO flow around the
cylinder obtained by the analytical model with those obtained by a
numerical solution of the full equation system, where the double layer is
fully resolved, we have noted significant discrepancies. These
discrepancies, which are described in the following, have become the
primary motivation for the study presented in this paper.

First, in the full double-layer resolving simulation we determined the
value $\Pmech^*(R_0) = \oint_{|\vec{r}|=R_0}
\hat{\vec{n}}\cdot\vec{\sigma}\cdot \vec{v}\,\dm a$ of the mechanical input
power, where $R_0$ is the radius of a cylinder surface placed co-axially
with the metallic cylinder. Then, as expected due to the electrical forces
acting on the net charge in the double layer, we found that $\Pmech^*(R_0)$
varied substantially as long as the integration cylinder surface was inside
the double layer. For $R_0 \approx a_c+6\lamD$ the mechanical input power
stabilized at a certain value. However, this value is significantly lower
than the analytical value, but the discrepancy decreased for decreasing
values of $\lamD$. Remarkably, even for a quite thin Debye layer, $\lamD =
0.01\:a_c$, the value of the full numerical simulation was about 40\% lower
than the analytical value. Clearly, the analytical model overestimates the
ICEO effect, and the double-layer width must be extremely thin before the
simple analytical model agrees well with the full model.

A partial explanation of the quantitative failure of the analytical slip
velocity model is the radial dependence of the tangential field
$E_\parallel$ combined with the spatial extent of the charge density
$\rho_\mathrm{el}$ of the double layer. In the Debye--H\"{u}ckel
approximation $E_\parallel$ and $\rho_\mathrm{el}$ around the metallic
cylinder of radius $a_c$ become
 \bsub
 \bal
 E_{\parallel}(r,\theta) &=
 E_0\left[1 + \frac{a_c^2}{r^2} - 2\frac{a_c}{r}\:
 \frac{K_1\Big(\frac{r}{\lamD}\Big)}{K_1\Big(\frac{a_c}{\lamD}\Big)}\right]
 \:\sin\theta,
 \\[2mm]
 \rho_\mathrm{el}(r,\theta) &= 2\:\frac{\epsilon E_0a_c}{\lamD^2}\:
 \frac{K_1\Big(\frac{r}{\lamD}\Big)}{K_1\Big(\frac{a_c}{\lamD}\Big)}\:\cos\theta,
 \eal
 \esub
where $K_1$ is the decaying modified Bessel function of order 1. The slowly
varying part of $E_\parallel$ is given by $E_0\big[1 +
(a_c/r)^2\big]\sin\theta$. For very thin double layers it is well
approximated by the $r$-independent expression $2E_0\sin\theta$, while for
wider double layers, the screening charges sample the decrease of
$E_\parallel$ as a function of the distance from the cylinder. Also
tangential hydrodynamic and osmotic pressure gradients developing in the
double layer may contribute to the lower ICEO strength when taking the
finite width of the double layer into account.

In this work we analyze quantitatively the impact of a finite Debye length
on the kinetic energy of the flow rolls generated by ICEO for three
different models: (\textit{i}) The full nonlinear electrokinetic model (FN)
with a fully resolved double layer, (\textit{ii}) the linear slip-velocity
model (LS), where electrostatics and hydrodynamics are completely
decoupled, and (\textit{iii}) a nonlinear slip-velocity model (NSL)
including the double layer charging through ohmic currents from the bulk
electrolyte and the surface conduction in the Debye layer. The latter two
models are only strictly valid for infinitely thin double layers, and we
emphasize that the aim of our analysis is to determine the errors
introduced by these models neglecting the finite width of the double layers
compared to the full nonlinear model resolving the double layer. We do not
seek to provide a more accurate description of the physics in terms of
extending the modeling by adding, say, the Stern layer (not present in the
model) or the steric effects of finite-sized ions (not taken into account).

\section{Model system}
\label{sec:Model sytem}
\begin{figure}[t]
 \centering
 \includegraphics[width=\columnwidth]{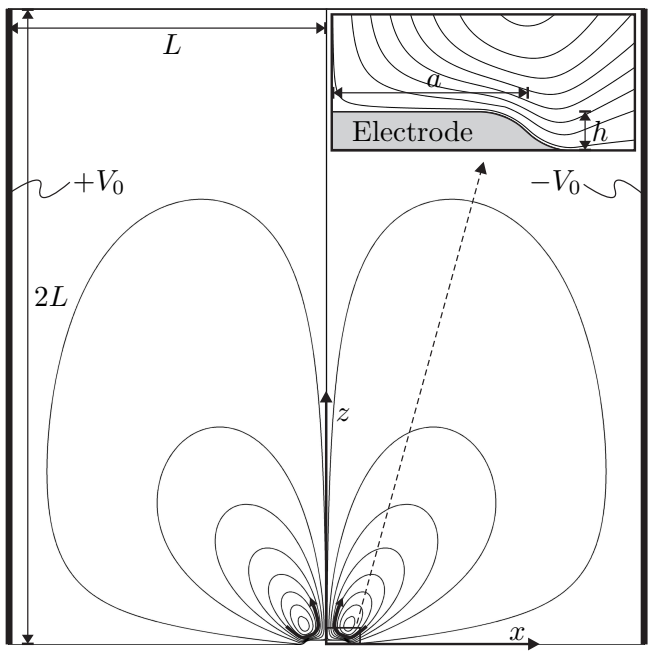}
\caption{\label{fig:ModelSystem} A sketch of the square $2L\times2L$
electrolytic microchamber in the $xz$-plane. The external voltage $\pm V_0$ is
applied to the two electrodes (thick black lines) at $x=\pm L$, respectively.
It induces two counter-rotating flow rolls (curved black arrows) by
electro-osmosis over the un-biased metallic center electrode of
length $2a$ and height $h$ placed at the bottom wall around
$(x,z)=(0,0)$. The spatial extent of the flow rolls is represented
by the streamline plot (thin black curves) drawn as equidistant
contours of the flow rate. The inset is a zoom-in on the right half,
$0<x<a$, of the un-biased center electrode and the nearby
streamlines.}
\end{figure}

To keep our analysis simple, we consider a single un-biased metallic
electrode in a uniform, external electric field. The electrode of
width $2a$ and height $h$ is placed at the bottom center, $-a< x <
a$ and $z=0$, of a square $2L\times 2L$ domain in the $xz$-plane
filled with an electrolyte, see Fig.~\ref{fig:ModelSystem}. The
system is unbounded and translational invariant in the perpendicular
$y$-direction. The uniform electric field, parallel to the surface
of the center electrode, is provided by biasing the driving
electrodes placed at the edges $x=\pm L$ with the DC voltages $\pm
V_0$, respectively. This anti-symmetry in the bias voltage ensures
that the constant potential of the center electrode is zero. A
double layer, or a Debye screening layer, is induced above the
center electrode, and an ICEO flow is generated consisting of two
counter-rotating flow rolls. Electric insulating walls at $z=0$ (for
$|x|>a$) and at $z=2L$ confine the domain in the $z$-direction. The
symmetry of the system around $x=0$ is exploited in the numerical
calculations.

\section{Full nonlinear model (FN)}
\label{sec:GoverningEqu}
We follow the usual continuum approach to the electrokinetic modeling of
the electrolytic microchamber and treat only steady-state problems. For
simplicity we consider a symmetric, binary electrolyte, where the positive
and negative ions with concentrations $c_+$ and $c_-$, respectively, have
the same diffusivity $D$ and charge number $Z$. Using the ideal gas model
for the ions, an ion is affected by the sum of an electrical and an osmotic
force given by $\vec{F}_\pm = \mp Ze\nablabf \phi - (\kBT/c_\pm)\: \nablabf
c_\pm$. Here $e$ is the elementary charge, $T$ is the absolute temperature
and $k_\textrm{B}$ is Boltzmann's constant. Assuming a complete force
balance between each ion and the surrounding electrolyte, the resulting
body force density $\vec{f}_ \mathrm{ion} = \sum_{i=\pm} c_i \vec{F}_i$,
appearing in the Navier--Stokes for the electrolyte due to the forces
acting on the ions, is
 \beq \label{eq:fIon}
 \vec{f}_ \mathrm{ion} =
 -Ze \big(c_+-c_-\big)\nablabf\phi - \kBT \nablabf\big(c_++c_-\big).
 \eeq
As the second term is a gradient, namely the gradient of the osmotic
pressure of the ions, it can in the Navier--Stokes equation be absorbed
into the pressure gradient $\nablabf p = \nablabf p_\mathrm{dyn} + \nablabf
p_\mathrm{os}$, which is the gradient of the sum of hydrodynamic pressure
and the osmotic pressure. Only the electric force is then kept as an
explicit body force.

\begin{figure*}[t]
 \centering
 \includegraphics[width=\textwidth]{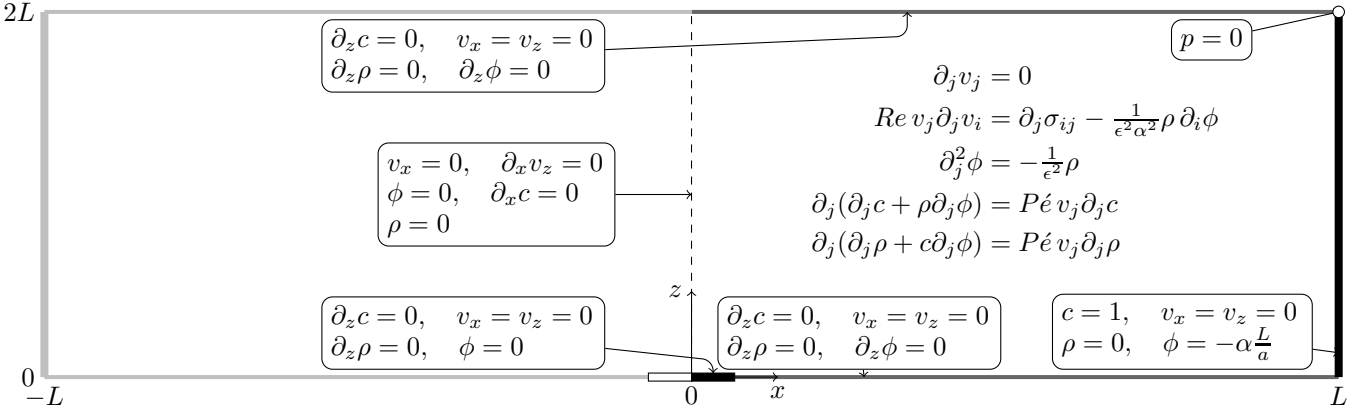}
\caption{\label{fig:Domain_linc} The governing equations (without box) and
boundary conditions (with boxes, arrows points to specific boundaries) in
dimensionless form (the tilde is omitted for clarity) for the entire
quadratic $2L\times2L$ domain (not shown in correct aspect ratio) bisected
into two symmetric halves. Only the right half ($x>0$) of the domain is
included in the simulations. The boundaries are the surface of the
un-biased center electrode (black rectangle), the solid insulating walls
(dark gray lines), the external electrode (black line), and the symmetry
line (dashed black line).}
\end{figure*}

\subsection{Bulk equations}
Neglecting bulk reactions in the electrolyte, the ionic transport is
governed by the particle conservation
 \begin{equation}\label{eq:Jcons_gov}
 \nablabf\cdot\vec{J}_\pm = 0,
 \end{equation}
where $\vec{J}_\pm$ is the flux density of the two ionic species.
Assuming the electrolytic solution to be dilute, the ion flux
densities are governed by the Nernst--Planck equation
 \begin{equation}\label{eq:J2_gov}
 \vec{J}_\pm = -D\bigg(
 \nablabf c_\pm +\frac{\pm Ze}{\kBT}c_\pm\nablabf\phi\bigg)
 + c_\pm\vec{v},
 \end{equation}
where the first term expresses ionic diffusion and the second term ionic
electromigration due to the electrostatic potential $\phi$. The last term
expresses the convective transport of ions by the fluid velocity field
$\vec{v}$.

The electrostatic potential is determined by the charge density
$\rhoe = Ze(c_+ - c_-)$ through Poisson's equation
\begin{equation}
\label{eq:phi1_gov} \nablabf\cdot(\varepsilon\nablabf\phi) = -\rhoe,
\end{equation}
where $\varepsilon$ is the fluid permittivity, which is assumed
constant. The fluid velocity field $\vec{v}$ and pressure field $p$
are governed the the continuity equation and the Navier--Stokes
equation for incompressible fluids,
 \bsub
 \begin{align}
 \label{eq:ContEq_gov}
 \nablabf\cdot\vec{v} &= 0,\\
 \label{eq:v1_gov}
 \rho_\textrm{m}(\vec{v}\cdot\nablabf)\vec{v}
 &= -\nablabf p + \eta \nabla^2\vec{v} - \rhoe\nablabf\phi,
 \end{align}
 \esub
where $\rho_\textrm{m}$ and $\eta$ are the fluid mass density and
viscosity, respectively, both assumed constant.

\subsection{Dimensionless form}
To simplify the numerical implementation, the governing equations are
rewritten in dimensionless form, as summarized in
Fig.~\ref{fig:Domain_linc}, using the characteristic parameters of the
system: The geometric half-length $a$ of the electrode, the ionic
concentration $c_0$ of the bulk electrolyte, and the thermal voltage
$\phi_0=\kBT/(Ze)$. The characteristic zeta-potential $\zeta$ of the center
electrode, i.e.\ its induced voltage, is given as the voltage drop along
half of the electrode, $\zeta = (a/L)V_0$, and we introduce the
dimensionless zeta-potential $\alpha$ as $\zeta \equiv \alpha \phi_0$, or
$\alpha = (aV_0)/(L\phi_0)$. The characteristic velocity $u_0$ is chosen as
the Helmholtz--Smoluchowski slip velocity induced by the local electric
field $E=\zeta/a$, and finally the pressure scale is set by the
characteristic microfluidic pressure scale $p_0=\eta u_0/a$. In summary,

 \begin{equation}\label{eq:u0Def}
 \phi_0 = \frac{\kBT}{Ze}, \quad
 u_0 = \frac{\varepsilon\zeta}{\eta}\frac{\zeta}{a}
 = \frac{\varepsilon\phi_0^2}{\eta
 a}\,\alpha^2, \quad
 p_0 = \frac{\eta u_0}{a}.
 \end{equation}
The new dimensionless variables (denoted by a tilde) thus become
 \begin{equation}
 \tilde{\vec{r}}=\frac{\vec{r}}{a}, \quad
 \tilde{c}_i=\frac{c_i}{c_0}, \quad
 \tilde{\phi}=\frac{\phi}{\phi_0}, \quad
 \tilde{\vec{v}}=\frac{\vec{v}}{u_0}, \quad
 \tilde{p}=\frac{p}{p_0}.
 \end{equation}

To exploit the symmetry of the system and the binary electrolyte, the
governing equations are reformulated in terms of the average ion
concentration $c\equiv (c_++c_-)/2$ and half the charge density $\rho\equiv
(c_+-c_-)/2$. Correspondingly, the average ion flux density
$\vec{J}_c=(\vec{J}_++\vec{J}_-)/2$ and half the current density
$\vec{J}_\rho=(\vec{J}_+-\vec{J}_-)/2$ are introduced. Thus, the resulting
full system of coupled nonlinear equations takes the following form for the
ionic fields
 \bsub
 \begin{align}
 \tilde{\nablabf}\cdot\tilde{\vec{J}}_{c} &=
  \tilde{\nablabf}\cdot\tilde{\vec{J}}_{\rho} = 0, \\
 \tilde{\vec{J}}_c&=-\tilde{\rho}\tilde{\nablabf}\tilde{\phi}
 -\tilde{\nablabf}\tilde{c} +\textit{P\'e}\:\tilde{c}\tilde{\vec{v}}, \\
 \tilde{\vec{J}}_\rho&=-\tilde{c}\tilde{\nablabf}\tilde{\phi}
 -\tilde{\nablabf}\tilde{\rho}
 +\textit{P\'e}\:\tilde{\rho}\tilde{\vec{v}},\\
 \textit{P\'e} &= \frac{u_0 a}{D},
 \end{align}
 \esub
while the electric potential obeys
 \begin{equation}
 \tilde{\nablabf}^2\tilde{\phi} = -\frac{1}{\epsilon^2}\tilde{\rho},
 \end{equation}
and finally the fluid fields satisfy
 \bsub
 \begin{align}
 \label{eq:ContEq}
 \tilde{\nablabf}\cdot\tilde{\vec{v}} &= 0,\\
 \textit{Re}\big(\tilde{\vec{v}} \cdot\tilde{\nablabf}\big)\tilde{\vec{v}}
 &= -\tilde{\nablabf}\tilde{p} +\tilde{\nablabf}^2\tilde{\vec{v}}
 -\frac{\tilde{\rho}}{\,\epsilon^2\,\alpha^2}\,  \tilde{\nablabf}\tilde{\phi}, \\
 \textit{Re} &= \frac{\rho u_0 a}{\eta}.
 \end{align}
 \esub
Here the small dimensionless parameter $\epsilon = \lamD/a$ has been
introduced, where $\lamD$ is the Debye length,
 \begin{equation}\label{eq:Debyelength}
 \epsilon = \frac{\lamD}{a} = \frac{1}{a}
 \sqrt{\frac{\varepsilon\kBT}{2(Ze)^2c_0}}.
 \end{equation}
We note that the dimensionless form of the osmotic force, the second term
in Eq.~(\ref{eq:fIon}), is $\tilde{\vec{f}}^\mathrm{os}_\mathrm{ion} =
-(1/\epsilon^2\alpha^2)\nablabf \tilde{c}$.

\begin{figure*}[t]
 \centering
 \includegraphics[scale=1]{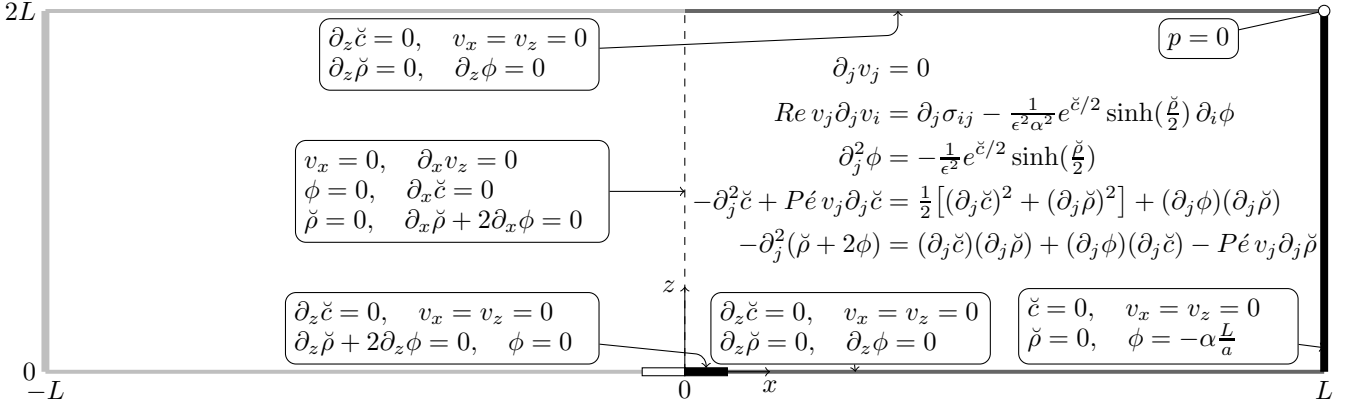}
\caption{\label{fig:Domain_logc} The governing equations (without box) and
boundary conditions (with boxes) in dimensionless form (the tilde is
omitted) using the logarithmic concentrations (denoted by a breve) of
Eq.~(\ref{eq:logc}). Otherwise the figure is identical to
Fig.~\ref{fig:Domain_linc}.}
\end{figure*}

\subsection{Boundary conditions}
We exploit the symmetry around $x=0$ and consider only the right
half ($0<x<L$) of the domain, see Fig.~\ref{fig:Domain_linc}. As
boundary conditions on the driving electrode we take both ion
concentrations to be constant and equal to the bulk charge neutral
concentration. Correspondingly, the charge density is set to zero.
Consequently, we ignore all dynamics taking place on the driving
electrode and simply treat it as an equipotential surface with the
value $V_0$. We set a no-slip condition for the fluid velocity, and
thus at $x=L$ we have
 \begin{equation}\label{eq:BC_full_drivel}
 \tilde{c}=1, \quad \tilde{\rho}=0, \quad
 \tilde{\phi}=\frac{V_0}{\phi_0}=\alpha\frac{L}{a}, \quad
 \tilde{\vec{v}}=\vec{0}.
 \end{equation}
On the symmetry axis ($x=0$) the potential and the charge density
must be zero due to the anti-symmetry of the applied potential.
Moreover, there is neither a fluid flux nor a net ion flux in the
normal direction and the shear stresses vanish. So at $x=0$ we have
 \bsub \label{eq:BC_full_sym}
 \begin{alignat}{3}
 \tilde{\phi} &=0, & \quad
 \hat{\vec{n}}\cdot\tilde{\vec{J}}_c &=0, & \quad
 \tilde{\rho} &=0,\\
 \hat{\vec{t}} \cdot\tilde{\vec{\sigma}} \cdot \hat{\vec{n}}&=0,
 &
 \hat{\vec{n}}\cdot\tilde{\vec{v}}&=0, &&
 \end{alignat}
 \esub
where the stress tensor is
$(\vec{\sigma})_{ik}=-p\delta_{ik}+\eta(\partial_iu_k+\partial_ku_i)$,
and $\hat{\vec{n}}$ and $\hat{\vec{t}}$ are the normal and
tangential unit vectors, respectively, which in 2D, contrary to 3D,
are uniquely defined. The constant potential on the un-biased
metallic electrode is zero due to symmetry, and on the electrode
surface we apply a no-slip condition on the fluid velocity and
no-current condition in the normal direction. So on the electrode
surface we have
 \begin{equation}\label{eq:BC_full_floatel}
 \hat{\vec{n}}\cdot\tilde{\vec{J}}_c=0, \quad
 \hat{\vec{n}}\cdot\tilde{\vec{J}}_\rho=0, \quad
 \tilde{\phi}=0, \quad \tilde{\vec{v}}=\vec{0}.
 \end{equation}
On the solid, insulating walls there are no fluxes in the normal
direction, the normal component of the electric field vanishes and
there are no-slip on the fluid velocity.
 \begin{equation}\label{eq:BC_full_wall}
 \hat{\vec{n}}\cdot\tilde{\vec{J}}_c=0, \quad
 \hat{\vec{n}}\cdot\tilde{\vec{J}}_\rho=0, \quad
 \hat{\vec{n}}\cdot\nablabf\tilde{\phi}=0,
 \quad \tilde{\vec{v}}=\vec{0}.
 \end{equation}

A complete overview of the governing equations and boundary
conditions is given in Fig.~\ref{fig:Domain_linc}.

\subsection{The strongly nonlinear regime}
At high values of the induced $\zeta$-potential, the concentrations
of counter- and co-ions acquire very large and very small values,
respectively, near the center electrode. Numerically this is
problematic. The concentration ratio becomes extremely large and the
vanishingly small concentration of co-ions is comparable to the
round-off error and may even become negative. However, this
numerical problem can be circumvented by working with the logarithms
(marked by a breve accent) of the concentration fields,
$\breve{c}_{\pm} = \log(c_\pm/c_0)$. By inserting
 \begin{equation} \label{eq:logc}
 c_\pm = c_0\,\exp\big(\breve{c}_\pm\big)
 \end{equation}
in the governing equations~\eqref{eq:J2_gov}, \eqref{eq:phi1_gov},
and~\eqref{eq:v1_gov}, a new equivalent set of governing equations
is derived. The symmetry is exploited by defining the symmetric
$\breve{c}=\breve{c}_++\breve{c}_-$ and antisymmetric
$\breve{\rho}=\breve{c}_+-\breve{c}_-$ combination of the
logarithmic fields and the corresponding formulation of the
governing equations is
 \bsub
 \begin{align}
 \tilde{\nabla}^2\breve{c} &= P\acute{e}\,
 \tilde{\vec{v}}\cdot\tilde{\nablabf}\breve{c}
 -\frac{(\tilde{\nablabf}\breve{c})^2
 \!+\!
 (\tilde{\nablabf}\breve{\rho})^2}{2}
 -\tilde{\nablabf}\tilde{\phi}\cdot\tilde{\nablabf}\breve{\rho},\\
 \tilde{\nabla}^2\big(\breve{\rho} + 2\tilde{\phi}\big)
 &= P\acute{e}\,\tilde{\vec{v}}\cdot\tilde{\nablabf}\breve{\rho}
 -\tilde{\nablabf}\breve{c} \cdot
 \tilde{\nablabf}\breve{\rho}
 -\tilde{\nablabf}\tilde{\phi}\cdot\tilde{\nablabf}\breve{\rho},\\
 \tilde{\nabla}^2\tilde{\phi} &= -\frac{1}{\epsilon^2}\,
 e^{\breve{c}/2}
 \sinh\left(\frac{\breve{\rho}}{2}\right),\\
 Re \big(\tilde{\vec{v}}\cdot\tilde{\nablabf}\big)\tilde{\vec{v}}
 &= -\tilde{\nablabf}\tilde{p}+\tilde{\nabla}^2\tilde{\vec{v}}
 -\frac{1}{\epsilon^2\alpha^2}\,
 e^{\breve{c}/2} \sinh\left(\frac{\breve{\rho}}{2}\right)
 \tilde{\nablabf}\tilde{\phi},
 \end{align}
 \esub
while the continuity equation remains the same as in Eq.~\eqref{eq:ContEq}.
The governing equations and boundary conditions for the logarithmic fields
(breve-notation) is summarized in Fig.~\ref{fig:Domain_logc}. This
transformation serves to help linearize solutions of the dependent
variables, $\breve{c}$ and $\breve{\rho}$, at the expense of introducing
more nonlinearity into the governing equations.

\section{Slip-velocity models}
\label{sec:ZeroWidthModels}
The numerical calculation of ICEO flows in microfluidic systems is
generally connected with computational limitations due to the large
difference of the inherent length scales. Typically, the Debye
length is much smaller than the geometric length scale, $\lamD \ll
a$, making it difficult to resolve both the dynamics of the Debye
layer and the entire microscale geometry with the available computer
capacity. Therefore, it is customary to use slip-velocity models,
where it is assumed that the electrodes are screened completely by
the Debye layer leaving the bulk electrolyte charge neutral. The
dynamics of the Debye layer is modeled separately and applied to the
bulk fluid velocity through an effective, so-called
Helmholtz--Smoluchowski slip velocity condition at the electrode
surface,
\begin{equation}\label{eq:HS_slipvel}
\vec{v}_\textrm{HS}=-\frac{\varepsilon}{\eta}\,\zeta\,\vec{E}_\parallel.
\end{equation}
where $\zeta$ is the zeta potential at the electrode surface, and
$\vec{E}_\parallel$ is the electric field parallel to the surface.
Regardless of the modeled dynamics in the double layer the
slip-velocity models are only strictly valid in the limit of
infinitely thin double layers $\lamD \ll a$.

\subsection{The linear slip-velocity model (LS)}
\label{sec:LinearModel}
The double-layer screening of the electrodes leaves the bulk
electrolyte charge neutral, and hence the governing equations only
include the potential $\phi$, the pressure field $p$ and the flow
velocity field $\vec{v}$. In dimensionless form they become,
 \bsub
 \label{eq:simple_slip_equ}
 \begin{align}
 \label{eq:Laplace}
 \tilde{\nablabf}^2\tilde{\phi} &= 0, \\
 \textit{Re}\big(\tilde{\vec{v}} \cdot
 \tilde{\nablabf}\big)\tilde{\vec{v}}
 &= -\tilde{\nablabf}\tilde{p}
 +\tilde{\nablabf}^2\tilde{\vec{v}}, \\
 \tilde{\nablabf}\cdot\tilde{\vec{v}} &= 0.
 \end{align}
 \esub
The electrostatic problem is solved independently of the
hydrodynamics, and the potential is used to calculate the effective
slip velocity applied to the fluid at the un-biased electrode
surface. The boundary conditions of the potential and fluid velocity
are equivalent to the conditions applied to the full non-linear
system, except at the surface of the un-biased electrode. Here, the
normal component of the electric field vanishes, and the effective
slip velocity of the fluid is calculated from the electrostatic
potential using $\zeta = -\phi$ and $\vec{E}_\parallel =
-\big[(\hat{\vec{t}}\cdot\tilde{\nablabf})
\tilde{\phi}\big]\,\hat{\vec{t}}$,
 \bsub
 \label{eq:simple_slip_BC}
 \begin{align}
 \label{eq:BC_simpphi}
 \hat{\vec{n}}\cdot\nablabf\tilde{\phi} &=0, \\
 \label{eq:BC_simpslip}
 \tilde{\vec{v}}_\textrm{HS}&= \frac{1}{\alpha^2}\,
 \tilde{\phi} \,\Big[(\hat{\vec{t}}\cdot\tilde{\nablabf})
 \tilde{\phi}\Big]\:\hat{\vec{t}}.
 \end{align}
 \esub

This represents the simplest possible, so-called linear slip-velocity
model; a model which is widely applied as a starting point for numerical
simulations of actual microfluidic systems \cite{Probstein1994,Bruus2008}.
In this simple model all the dynamics of the double layer has been
neglected, an assumption known to be problematic when the voltage across
the electrode exceeds the thermal voltage.

\subsection{The nonlinear slip-velocity model (NLS)}
\label{sec:NonlinearModel}
The linear slip-velocity model can be improved by taking into account the
nonlinear charge dynamics of the double layer. This is done in the
so-called nonlinear slip-velocity model, where, although still treated as
being infinitely thin, the double layer has a non-trivial charge dynamics
with currents from the bulk in the normal direction and currents flowing
tangential to the electrode inside the double layer. For simplicity we
assume in the present nonlinear model that the neutral salt concentration
$c_0$ is uniform. This assumption breaks down at high zeta potentials,
where surface transport of ionic species can set up gradients in the salt
concentrations leading to chemi-osmotic flow. In future more complete
studies of double layer charge dynamics these effects should be included.

The charging of the double layer by the ohmic bulk current is assumed to
happen in quasi-equilibrium characterized by a nonlinear differential
capacitance $C_\textrm{dl}$ given by the Gouy--Chapmann model,
$C_\textrm{dl}=\varepsilon\cosh[ze\zeta/(2\kBT)]/\lamD$, which in the the
low-voltage, linear Debye--H\"{u}ckel regime reduces to
$C_\textrm{dl}=\varepsilon/\lamD$. Ignoring the Stern layer, the
zeta-potential is directly proportional to the bulk potential right outside
the double layer, $\zeta = -\phi$.

The current along the electrode inside the Debye layer is described
by a 2D surface conductance $\sigma_s$, which for a binary,
symmetric electrolyte is given by \cite{Dukhin1993}
\begin{equation}
    \sigma_s = 4\lamD\sigma(1+m)\sinh^2
    \left(\frac{Ze\zeta}{4\kBT}\right),
\end{equation}
where $\sigma$ is the bulk 3D conductivity and
\begin{equation}
    m = 2\frac{\varepsilon}{\eta D}\left(\frac{\kBT}{Ze}\right)^2
\end{equation}
is a dimensionless parameter indicating the relative contribution of
electroosmosis to surface conduction. In steady state the conservation of
charge then yields \cite{Soni2009}
\begin{equation}
 \label{eq:EffectiveBoundaryEqu}
 0 = \hat{\vec{n}}\cdot(\sigma\nablabf\phi) +
 \nablabf_s\cdot
 \big[\sigma_s\nablabf_s\phi\big],
\end{equation}
where the operator $\nablabf_s = \hat{\vec{t}}(\hat{\vec{t}}\cdot\nablabf)$
is the gradient in the tangential direction of the surface.

Given the length scale $a$ of the electrode, the strength of the
surface conductance can by characterized by the dimensionless Dukhin
number \emph{Du} defined by
\begin{equation}
    Du = \frac{\sigma_s}{a\sigma} =
    \frac{4\lamD}{a}(1+m)\sinh^2\left(\frac{Ze\zeta}{\kBT}\right).
\end{equation}
Conservation of charge then takes the dimensionless form
\begin{equation}
 \label{eq:DimLessBoundaryEqu}
 0 = \hat{\vec{n}}\cdot(\tilde{\nablabf}\tilde{\phi}) +
 \tilde{\nablabf}_s\cdot
 \big[Du\tilde{\nablabf}_s\cdot\tilde{\phi}\big],
\end{equation}
and this effective boundary condition for the potential on the flat
electrode constitutes a 1D partial differential equation and as such
needs accompanying boundary conditions. As a boundary condition the
surface flux is assumed to be zero at the edges of the electrode,
\begin{equation}
  \sigma_s(\hat{\vec{t}}\cdot\nablabf)\phi\big|_{x=\pm a} = 0,
\end{equation}
which is well suited for the weak formulation we employ in our
numerical simulation as seen in Eq.~\eqref{eq:nofluxint}.

\section{Numerics in COMSOL}
\label{sec:COMSOL}
The numerical calculations are performed using the commercial
finite-element-method software COMSOL with second-order Lagrange
elements for all the fields except the pressure, for which
first-order elements suffices. We have applied the so-called weak
formulation mainly to be able to control the coupling between the
bulk equations and the boundary constraints, such as
Eqs.~(\ref{eq:BC_simpslip}) and~(\ref{eq:EffectiveBoundaryEqu}), in
the implementation of the slip-velocity models in script form.

The Helmholtz--Smoluchowski slip condition poses a numerical
challenge because it is a Dirichlet condition including not one, but
up to three variables, for which we want a one-directional coupling
from the electrostatic field $\phi$ to the hydrodynamic fields
$\vec{v}$ and $p$. We use the weak formulation to unambiguously
enforce the boundary condition with the explicit introduction of the
required hydrodynamic reaction force $\vec{f}$ on the un-biased
electrode
 \begin{equation}
 \vec{f} = \vec{\sigma}\cdot\hat{\vec{n}}.
 \end{equation}
The $x$ and $z$ components of Navier--Stokes equation are multiplied
with test functions $u_x$ and $u_z$, respectively, and subsequently
integrated over the whole domain $\Omega$. Partial integration is
then used to move the stress tensor contribution to the boundaries
$\partial \Omega$,
 \begin{equation}
 0 = \int_{\partial\Omega} u_i\sigma_{ij}n_j\textrm{d}s
 - \int_\Omega \big[(\partial_ju_i)\sigma_{ij}
 + u_iB_i\big] \textrm{d}a,
 \end{equation}
where $B_i=Re\left(v_j\partial_j\right)v_i +
\rho(\partial_i\phi)/(\epsilon^2\alpha^2)$. The boundary integral on
the un-biased electrode $\partial\Omega_\textrm{ue}$ is rewritten as
 \begin{equation}
 \int_{\partial\Omega_\textrm{ue}} u_i\sigma_{ij}n_j\textrm{d}s
 = \int_{\partial\Omega_\textrm{ue}} \big[
 u_if_i+F_i(v_i-v_{\textrm{HS},i})\big]\textrm{d}s,
 \end{equation}
where $F_i$ are the test functions belonging to the components $f_i$
of the reaction force $\vec{f}$. These test functions are used to
enforce the Helmholtz--Smoluchowski slip boundary condition
consistently. This formulation is used for both slip-velocity
models.

In the nonlinear slip-velocity model the Laplace
equation~(\ref{eq:Laplace}) is multiplied with the electrostatic test
function $\Phi$ and partially integrated to get a boundary term and a bulk
term
 \begin{equation}
 0 = \int_{\partial\Omega}
 \Phi\left(\partial_i\phi\right)n_i\textrm{d}s - \int_\Omega
 \left(\partial_i\Phi\right)\left(\partial_i\phi\right)\textrm{d}a.
 \end{equation}
The boundary integration term on the electrode is simplified by
substitution of Eq.~\eqref{eq:EffectiveBoundaryEqu} which results in
 \begin{equation}
 \int_{\partial\Omega_\textrm{ue}}
 \Phi\left(\partial_i\phi\right)n_i\textrm{d}s =
 -\int_{\partial\Omega_\textrm{ue}}
 \Phi\left[\hat{t}_i\partial_i
 \left(\textit{Du}\,\hat{t}_j\partial_j\phi\right)\right]\textrm{d}s.
 \end{equation}
Again, the resulting boundary integral is partially integrated,
which gives us explicit access to the end-points of the un-biased
electrode. This is necessary for applying the boundary conditions on
this 1D electrode,
 \begin{align}\label{eq:nofluxint}
 &\int_{\partial\Omega_\textrm{ue}}\!\!
 \Phi\Big[\hat{t}_i\partial_i(
 \textit{Du}\,\hat{t}_j\partial_j\phi)\Big]\textrm{d}s \nonumber \\
 &\quad = \Big[\Phi
 \textit{Du}\,(\hat{t}_i\partial_i\phi)\Big]_{x=-a}^{x=+a}
 - \int_{\partial\Omega_\textrm{ue}}\!\!
 (\hat{t}_i\partial_i\Phi)
 \textit{Du}\,(\hat{t}_j\partial_j\phi)\,\textrm{d}s,
 \end{align}
The no-flux boundary condition can be explicitly included with this
formulation. Note that in both slip-velocity models the
zeta-potential is given by the potential just outside the Debye
layer, $\zeta = -\phi$, and it is therefore not necessary to include
it as a separate variable.

The accuracy and the mesh dependence of the simulation as been
investigated as follows. The comparison between the three models
quantifies relative differences of orders down to $10^{-3}$, and the
convergence of the numerical results is ensured in the following
way. COMSOL has a build-in adaptive mesh generation technique that
is able to refine a given mesh so as to minimize the error in the
solution. The adaptive mesh generator increases the mesh density in
the immediate region around the electrode to capture the dynamics of
the ICEO in the most optimal way under the constraint of a maximum
number of degrees of freedom (DOFs). For a given set of physical
parameters, the problem is solved each time increasing the number of
DOFs and comparing consecutive solutions. As a convergence criterium
we demand that the standard deviation of the kinetic energy relative
to the mean value should be less than a given threshold value
typically chosen to be around $10^{-5}$. All of the simulations
ended with more than $10^6$ DOFs, and the ICEO flow is therefore
sufficiently resolved even for the thinnest double layers in our
study for which $\epsilon = 10^{-4}$.

\section{Results}
\label{sec:Results}

Our comparison of the three numerical models is primarily focused on
variations of the three dimensionless parameters $\epsilon$,
$\alpha$, and $\beta$ relating to the Debye length $\lamD$, the
applied voltage $V_0$, and the height $h$ of the electrode,
respectively,
 \begin{equation}
 \epsilon = \frac{\lamD}{a}, \quad
 \alpha = \frac{aV_0}{L\phi_0}, \quad
 \beta = \frac{h}{a}.
\end{equation}

As mentioned in Sec.~\ref{sec:Introduction}, the strength of the
generated ICEO flow can be measured as the mechanical power input
$\Pmech$ exerted on the electrolyte by the slip-velocity just
outside the Debye layer or equivalently by the kinetic energy
dissipation $\Pkin$ in the bulk of the electrolyte. However, both
these methods suffers from numerical inaccuracies due to the
dependence of both the position of the integration path and of the
less accurately determined velocity gradients in the stress tensor
$\vec{\sigma}$. To obtain a numerically more stable and accurate
measure, we have chosen in the following analysis to characterize
the strength of the ICEO flow by the kinetic energy $\Ekin$ of the
induced flow field $\vec{v}$,
 \begin{equation} \label{Ekin_def}
 \Ekin = \mbox{$\frac{1}{2}$} \rho_\textrm{m}
 \int_\Omega \!v^2\: \dm x\, \dm z,
 \end{equation}
which depends on the velocity field and not its gradients, and which
furthermore is a bulk integral of good numerical stability.

\begin{figure}[!t]
 \centering
 \includegraphics[width=1\columnwidth]{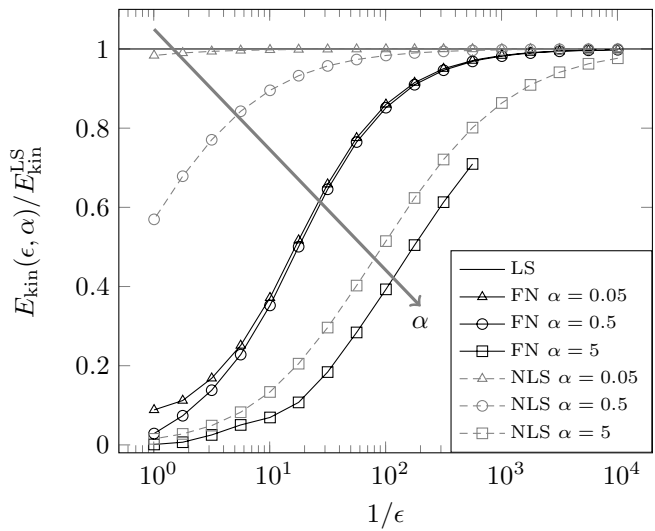}
\caption{\label{fig:EkinFullAdvEps} The total induced kinetic energy
$\EkinNlin$ (gray dashed) and $\EkinFull$ (black) for the nonlinear
slip-velocity model and the full model, respectively, relative to
$\EkinLin$ (horizontal black line) of the linear slip-velocity model
as a function of dimensionless inverse Debye length $1/\epsilon$.
Each are shown for three values of the dimensionless applied voltage
$\alpha = 0.05, 0.5$ and~5. The value of $\epsilon$ decreases from 1
to $10^{-4}$ going from left to right.}
\end{figure}

\subsection{Zero height of the un-biased center electrode}
\label{sec:ZeroHeight}

We assume the height $h$ of the un-biased center electrode to be zero,
i.e.\ $\beta = 0$, while varying the Debye length and the applied voltage
through the parameters $\epsilon$ and $\alpha$. We note that the linear
slip-velocity model Eqs.~(\ref{eq:simple_slip_equ})
and~(\ref{eq:simple_slip_BC}) is independent of the dimensionless Debye
length $\epsilon$. It is therefore natural to use the kinetic energy
$\EkinLin$ of this model as a normalization factor.

In the lin-log plot of Fig.~\ref{fig:EkinFullAdvEps} we show the kinetic
energy $\EkinNlin$ and $\EkinFull$ normalized by $\EkinLin$ as a function
of the inverse Debye length $1/\epsilon$ for three different values of the
applied voltage, $\alpha = 0.05, 0.5$ and 5, ranging from the linear to the
strongly nonlinear voltage regime.

We first note that in the limit of vanishing Debye length (to the
right in the graph) all models converge towards the same value for
all values of the applied voltage $\alpha$. For small values of
$\alpha$ the advanced slip-velocity model $\EkinNlin$ is fairly
close to the linear slip-velocity model $\EkinLin$, but as $\alpha$
increases, it requires smaller and smaller values of $\epsilon$ to
obtain the same results in the two models. In the linear regime
$\alpha=0.05$ a deviation less than 5\% is obtained already for
$\epsilon < 1$. In the nonlinear regime $\alpha = 0.5$ the same
deviation requires $\epsilon < 10^{-2}$, while in the strongly
nonlinear regime $\epsilon < 10^{-4}$ is needed to obtain a
deviation lower than 5\%.

In contrast, it is noted how the more realistic full model
$\EkinFull$ deviates strongly from $\EkinLin$ for most of the
displayed values of $\epsilon$ and $\alpha$. To obtain a relative
deviation less than 5\% in the linear ($\alpha = 0.05$) and
nonlinear ($\alpha = 0.5$) regimes, a minute Debye length of
$\epsilon < 10^{-3}$ is required, and in the strongly nonlinear
regime the 5\% level it not reached at all.

The deviations are surprisingly large. The Debye length in typical
electrokinetic experiments is $\lamD = 30$~nm. For a value of $\epsilon =
0.01$ this corresponds to an electrode of width $2\times3~\SImum =
6~\SImum$, comparable to those used in
Refs.~\cite{Studer2004,Cahill2004,Gregersen2007}. In
Fig.~\ref{fig:EkinFullAdvEps} we see that for $\alpha = 5$, corresponding
to a moderate voltage drop of 0.26~V across the electrode, the linear
slip-velocity model overestimates the ICEO strength by a factor 1/0.4 =
2.5. The nonlinear slip-model does a better job. For the same parameters it
only overestimates the ICEO strength by a factor 0.5/0.4 = 1.2.

\begin{figure}[!t]
 \centering
 \includegraphics[width=1\columnwidth]{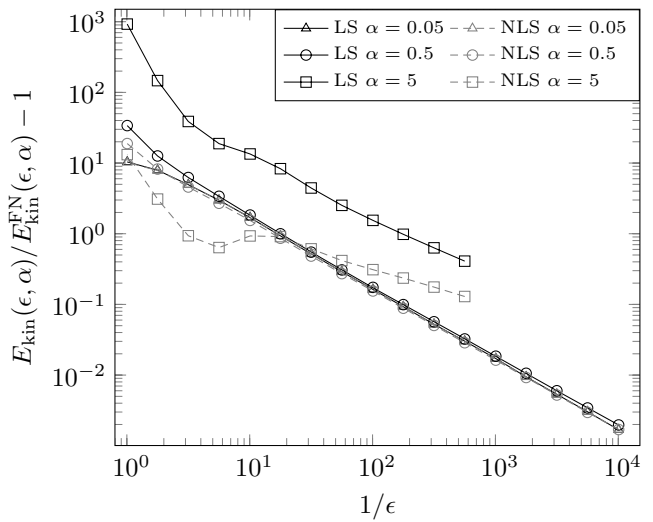}
\caption{\label{fig:DevEps} The difference between the induced
kinetic energies $\EkinLin$ and $\EkinNlin$ of the linear and
nonlinear slip-velocity models, respectively, relative to the full
model $\EkinFull$ as a function of the inverse Debye length
$1/\epsilon$. for three different applied voltages $\alpha = 0.05,
0.5, 5$. }
\end{figure}
\begin{figure}[!t]
 \centering
 \includegraphics[width=1\columnwidth]{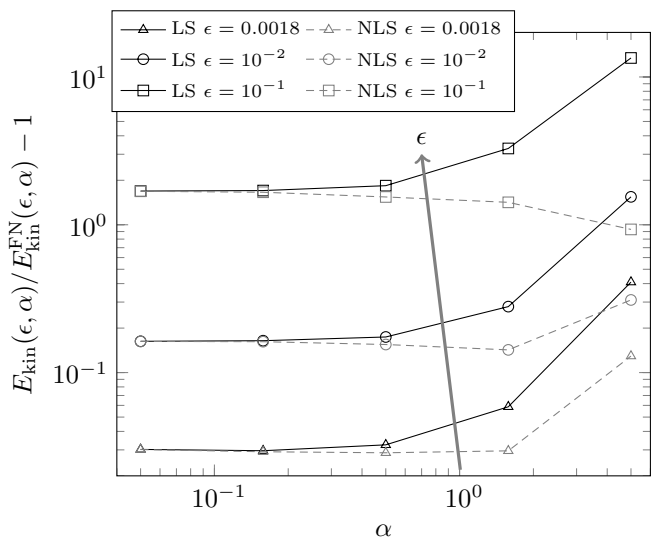}
\caption{\label{fig:DevAlpha} The difference between the induced
kinetic energies $\EkinLin$ and $\EkinNlin$ of the linear and
nonlinear slip-velocity models, respectively, relative to the full
model $\EkinFull$ as a function of the voltage bias $\alpha$ for
three different Debye layer thicknesses $\epsilon =
1.8\times10^{-3}, 10^{-2}, 10^{-1}$. }
\end{figure}

For more detailed comparisons between the three models the data of
Fig.~\ref{fig:EkinFullAdvEps} are plotted in a different way in
Fig.~\ref{fig:DevEps}. Here the overestimates $(\EkinLin/\EkinFull) -1$ and
$(\EkinNlin/\EkinFull) -1$ of the two slip-velocity models relative to the
more correct full model are plotted in a log-log plot as a function of the
inverse Debye length $1/\epsilon$ for three different values of the applied
voltage. It is clearly seen how the relative deviation decreases
proportional to $\epsilon$ as $\epsilon$ approaches zero.

Finally, in Fig.~\ref{fig:DevAlpha} the relative deviations
$(\EkinLin/\EkinFull) -1$ and $(\EkinNlin/\EkinFull) -1$ are plotted versus
the voltage $\alpha$ in a log-log plot. For any value of the applied
voltage $\alpha$, both slip-velocity models overestimates by more than
100\% for large Debye lengths $\epsilon = 10^{-1}$ and by more than 10\%
for $\epsilon = 10^{-2}$. For the minute Debye length $\lamD = 1.8\times
10^{-3}$ the overestimates are about 3\% in the linear and weakly nonlinear
regime $\alpha < 1$, however, as we enter the strongly nonlinear regime
with $\alpha=5$ the overestimation increases to a level above 10\%.

\subsection{Finite height of the un-biased electrode}
\label{sec:FiniteHeight}

Compared to the full numerical model, the slip-velocity models are
convenient to use, but even for small Debye lengths, say $\lamD = 0.01 a$,
they are prone to significant quantitative errors as shown above.
Similarly, it is of relevance to study how the height of the un-biased
electrode influences the strength of the ICEO flow rolls. In experiments
the thinnest electrodes are made by evaporation techniques. The resulting
electrode heights are of the order 50~nm $-$ 200~nm, which relative to the
typical electrode widths $a \approx 5~\SImum$ results in dimensionless
heights $10^{-3} < \beta < 10^{-1}$.

\begin{figure}[b]
 \centering
 \includegraphics[width=1\columnwidth]{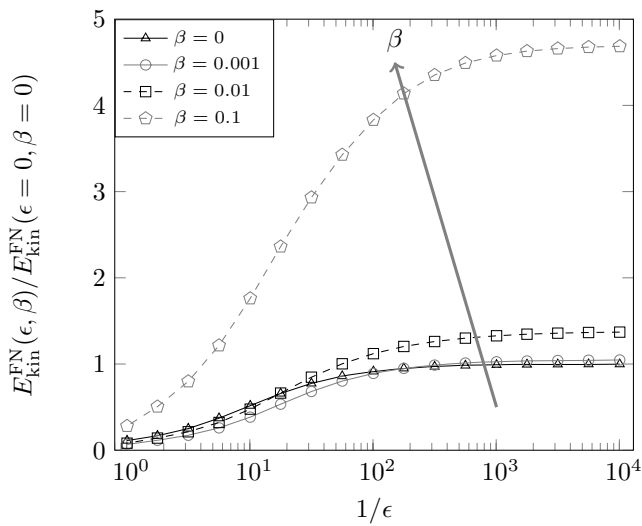}
\caption{\label{fig:EkinBeta} The difference between the induced
kinetic energies $\EkinFull(\epsilon,\beta)$ of the full model at
finite Debye length and electrode height relative to the full model
$\EkinFull(0,0)$ at zero Debye length and zero electrode height as a
function of the inverse Debye length $1/\epsilon$ for four electrode
heights $\beta = 0, 10^{-3}, 10^{-2}, 10^{-1}$.}
\end{figure}

In Fig.~\ref{fig:EkinBeta} is shown the results for the numerical
calculation of the kinetic energy $\EkinFull(\epsilon,\beta)$ using
the full numerical model. The dependence on the kinetic energy of
the dimensionless Debye length $\epsilon = \lamD/a$ and the
dimensionless electrode height $\beta = h/a$ is measured relative to
the value $\EkinFull(\epsilon,\beta)$ of the infinitely small Debye
length for an electrode of zero height.
For small values of the height no major deviations are seen. The curve for
$\beta = 0$ and  $\beta = 0.001$ are close. As the height is increased to
$\beta = 10^{-2}$ we note that the strength of the ICEO is increased by
20\%$-$25\% as $\beta > \epsilon$. This tendency is even stronger
pronounced for the higher electrode $\beta = 10^{-1}$. Here the ICEO
strength is increased by approximately 400\% for a large range of Debye
lengths.

\subsection{Thermodynamic efficiency of the ICEO system}
\label{sec:Efficiency}

Conventional electro-osmosis is known to have a low thermodynamic
efficiency defined as the delivered mechanical pumping power relative to
the total power delivered by the driving voltage. Typical efficiencies are
of the order of 1\%~\cite{Laser2004}, while in special cases an efficiency
of 5.6\% have been reported~\cite{Reichmuth2003}. In the following we
provide estimates and numerical calculations of the corresponding
thermodynamic efficiency of the ICEO system.

The applied voltage drop $2V_0 = 2E_0L$ across the system in the
$x$-direction is written as the average electrical field $E_0$ times
the length $2L$, while the electrical current is given by
$I=WH\sigma E_0$, where $W$ and $H$ is the width and height in the
$y$- and $z$-direction, respectively, and $\sigma =
D\varepsilon/\lamD^2 = \varepsilon/\tau_D$ is the conductivity
written in terms of the Debye time $\tau_D = \lamD^2/D$. The total
power consumption to run the ICEO system is thus
 \beq \label{eq:Ptot}
 \Ptot = 2V_0 \times I = \frac{4}{\tauD}\:
 \Big(\frac{1}{2}\varepsilon E_0^2\Big) LWH.
 \eeq
This expression can be interpreted as the total energy,
$\frac{1}{2}\varepsilon E_0^2\times LWH$, stored in the average
electrical field of the system with volume $LWH$ multiplied by the
characteristic electrokinetic rate $4/\tauD$.

The velocity-gradient part of the hydrodynamic stress tensor is denoted
$\tilde{\vec{\sigma}}$, i.e.\ $(\tilde{\vec{\sigma}})_{ij} =
\eta(\partial_iv_j + \partial_jv_i)$. In terms of $\tilde{\vec{\sigma}}$,
the kinetic energy dissipation $\Pkin$ necessary to sustain the
steady-state flow rolls is given by $\Pkin = \frac{W}{2\eta} \int_0^L \dm
x\:\int_0^H\dm z\: \mathrm{Tr}(\tilde{\vec{\sigma}}^2)$. In the following
estimate we work in the Debye--H\"{u}ckel limit for an electrode of length
$2a$, where the induced zeta potential is given by $\zeta_\mathrm{ind} =
aE_0$ and the radius of each flow roll is approximately $a$. In this limit
the electro-osmotic slip velocity $\ueo$ and the typical size of the
velocity gradient $|\partial_i v_j|$ are
 \bsub
 \label{eq:ueogradu}
 \begin{align}
 \label{eq:ueo}
 \ueo & = \frac{\varepsilon\zeta_\mathrm{ind}}{\eta}\:E_0
 = \frac{\varepsilon a}{\eta}\:E_0^2,\\
 \label{eq:gradu}
 |\partial_i v_j| &\approx \frac{\ueo}{a}
 = \frac{\varepsilon}{\eta}\:E_0^2.
 \end{align}
 \esub
Thus, since the typical area covered by each flow roll is $\pi a^2$
, we obtain the following estimate of $\Pkin$,
 \beq \label{eq:Pkin}
 \Pkin \approx 2\frac{W}{2\eta}\:
 \pi a^2\: 4\bigg[\eta\frac{\ueo}{a}\bigg]^2 =
 8\:\frac{\varepsilon E_0^2}{\eta}\:
 \Big(\frac{1}{2}\varepsilon E_0\Big)^2 \pi a^2 W.
 \eeq
Here the power dissipation can be interpreted as the energy of the
electrical field in the volume $\pi a^2 W$ occupied by each flow
roll multiplied by an ICEO rate given by the electric energy density
$\varepsilon E_0^2$ divided by the rate of viscous energy
dissipation per volume given by $\eta$.

The thermodynamic efficiency can now be calculated as the ratio
$\Pkin/\Ptot$ using Eqs.~(\ref{eq:Ptot}) and~(\ref{eq:Pkin}),
 \beq \label{eq:ThermoEff}
 \frac{\Pkin}{\Ptot} \approx
 \frac{2\pi a^2}{LH}\: \frac{\varepsilon E_0^2}{\eta/\tauD}
 \approx 2.4 \times 10^{-8}.
 \eeq
This efficiency is the product of the ratio between the volumes of the flow
rolls and the entire volume multiplied and the ratio of the electric energy
density in the viscous energy density $\eta/\tauD$. The value is found
using $L=H=15a=0.15$~mm, $E_0 = 2.5$~kV/m, and $\lamD = 20$~nm, which is in
agreement with the conventional efficiencies for conventional
electro-osmotic systems quoted above.

\section{Conclusion\label{sec:Conclusion}}

We have shown that the ICEO velocities calculated using the simple
zero-width models significantly overestimates those calculated in more
realistic models taking the finite size of the Debye screening length into
account. This may provide a partial explanation of the observed
quantitative discrepancy between observed and calculated ICEO velocities.
The discrepancy increases substantially for increasing $\epsilon$, i.e.\ in
nanofluidic systems.

Even larger deviations of the ICEO strength is calculated in the
full numerical model when a small, but finite height of the
un-biased electrode is taken into account.

A partial explanation of the quantitative failure of the analytical slip
velocity model is the decrease of the tangential electric field as a
function of the distance to the surface of the polarized ICEO object
combined with the spatial extent of the charge density of the double layer.
Also tangential hydrodynamic and osmotic pressure gradients developing
inside the double layer may contribute to the lowering ICEO strength when
taking the finite width of the double layer into account. The latter may be
related to the modification of the classical Helmholtz--Smoluchowski
expression of the slip-velocity obtained by adding a term proportional to
the gradient of the salt concentration $c$~\cite{Khair2008b}.

Our work shows that for systems with a small, but non-zero Debye length of
0.001 to 0.01 times the size of the electrode, and even when the
Debye-H\"{u}ckel approximation is valid, a poor quantitative agreement
between experiments and model calculations must be expected when applying
the linear slip-velocity model based on a zero Debye-length. It is advised
to employ the full numerical model of ICEO, when comparing simulations with
experiments.

\section{Acknowledgements}

We thank Sumita Pennathur and Martin Bazant for illuminating discussions,
and we are particularly grateful to Todd Squires for a number of valuable
comments and suggestions. This work is supported in part by the Institute
for Collaborative Biotechnologies through contract no.\ W911NF-09-D-0001
from the U.S. Army Research Office.  The content of the information herein
does not necessarily reflect the position or policy of the Government and
no official endorsement should be inferred.


\begin{thebibliography}{10}



 \bibitem{Dukhin1993}
 S.S.~Dukhin,
 Adv. Colloid Interface Sci. \textbf{44}, 1 (1993).

 \bibitem{Murtsovkin1996}
 V.A.~Murtsovkin,
 Colloid J. \textbf{58}, 341 (1996).

 \bibitem{Gonzales2000}
 A.~Gonzalez, A.~Ramos, N.G.~Green, A.~Castellanos and H.~Morgan,
 Phys. Rev. E \textbf{61}(4), 4019 (2000).

 \bibitem{Green2002}
 N.G.~Green, A.~Ramos, A.~Gonzalez, H.~Morgan and A.~Castellanos,
 Phys. Rev. E \textbf{66}, 026305 (2002).

 \bibitem{Ajdari2000}
 A.~Ajdari,
 Phys. Rev. E \textbf{61}, R45 (2000).

 \bibitem{Brown2000}
 A.B.D.~Brown, C.G.~Smith and A.R.~Rennie, Phys. Rev. E
 \textbf{63}, 016305 (2000).

 \bibitem{Studer2004}
 V.~Studer, A.~P\'epin, Y.~Chen and A.~Ajdari,
 The Analyst \textbf{129}, 944 (2004).

 \bibitem{Lastochkin2004}
 D.~Lastochkin, R.~Zhou, P.~Whang, Y.~Ben and H.-C.~Chang, J. Appl.
 Phys. \textbf{96}, 1730 (2004).

 \bibitem{Debesset2004}
 S.~Debesset, C.J.~Hayden, C.~Dalton, J.C.T.~Eijkel and A.~Manz,
 Lab Chip \textbf{4}, 396 (2004).

 \bibitem{Cahill2004}
 B.P.~Cahill, L.J.~Heyderman, J.~Gobrecht and A.~Stemmer,
 Phys. Rev. E \textbf{70}, 036305 (2004).

 \bibitem{Gregersen2007}
 M.M.~Gregersen, L.H.~Olesen, A.~Brask, M.F.~Hansen, and H.~Bruus,
 Phys. Rev. E \textbf{76} 056305 (2007).

 \bibitem{Olesen2006}
 L.H.~Olesen, H.~Bruus and A.~Ajdari,
 Phys. Rev. E \textbf{73}, 056313 (2006).

 \bibitem{Squires2004}
 T.M.~Squires, and M.Z.~Bazant, J. Fluid Mech. \textbf{509}, 217 (2004).

 \bibitem{Levitan2005}
 J.A.~Levitan, S.~Devasenathipathy, V.~Studer, Y.~Ben,
 T.~Thorsen, T.M.~Squires, and M.Z.~Bazant,
 Coll. Surf. A \textbf{267}, 122 (2005).

 \bibitem{Squires2006}
 T.M.~Squires, and M.Z.~Bazant, J. Fluid Mech. \textbf{560}, 65 (2006).

 \bibitem{Harnett2008}
 C.K.~Harnett, J.~Templeton, K.~Dunphy-Guzman, Y.M.~Senousy, and M.P.~Kanouff,
 Lab Chip \textbf{8}, 565 (2008).

 \bibitem{Khair2008}
 A.S.~Khair and T.M.~Squires,
 J. Fluid Mech. \textbf{615}, 323 (2008).

 \bibitem{Gregersen2009}
 M.M.~Gregersen, F.~Okkels, M.Z.~Bazant, and H.~Bruus,
 New J. Phys. (submitted, 2008),\\ http:/$\!$/arxiv.org/abs/0901.1788

 \bibitem{Soni2007}
 G.~Soni, T.M.~Squires, and C.D.~Meinhart,
 in \emph{Proceedings of 2007 ASME International Mechanical Engineering
 Congress and Exposition}, 2007.

 \bibitem{Probstein1994}
 R.F.~Probstein,
 \emph{Physicochemical hydrodynamics}, (John Wiley \& Sons, New York, 1994).

 \bibitem{Bruus2008}
 H.~Bruus,
 \emph{Theoretical Microfluidics}, (Oxford University Press, Oxford, 2008).

 \bibitem{Dukhin2001}
 S.S.~Dukhin, R.~Zimmermann, and C.~Werner,
 Coll. Surf. A \textbf{195}, 103 (2001).

 \bibitem{Lyklema2001}
 J.~Lyklema,
 J. Phys.: Condens. Matter \textbf{13}, 5027 (2001).

 \bibitem{Chu2007}
 K.T.~Chu, and M.Z.~Bazant,
 J. Colloid. Interf. Sci. \textbf{315}, 319 (2007).

 \bibitem{Soni2009}
 G.~Soni, M.B.~Andersen, H.~Bruus, T.~Squires, C.~Meinhart,
 Phys. Rev. E (in preparation, 2009)

 \bibitem{Laser2004}
 D.J.~Laser and J.G.~Santiago
 J. Micromech. Microeng. \textbf{14}, R1 (2004).

 \bibitem{Reichmuth2003}
 D.S.~Reichmuth, G.S.~Chirica, and B.J.~Kirby
 Sens. Actuators B \textbf{92}, 37 (2003).

  \bibitem{Khair2008b}
  A.S.~Khair and T.M.~Squires,
  Phys. Fluids \textbf{20}, 087102 (2008)

\end{thebibliography}
\end{document}